\documentclass[12pt]{article}
\usepackage{graphicx}
\usepackage{amssymb}
\usepackage{amsmath}
\usepackage{cite}
\usepackage{geometry}               
\newcommand {\slsh} [1] {\not{\hbox{\kern-2pt${#1}$}}}

\newcommand {\beq} {\begin{equation}}
\newcommand {\eeq} {\end{equation}}
  \newcommand {\ber}{\begin{eqnarray*}}
  \newcommand {\eer} {\end{eqnarray*}}
\newcommand {\beqn}{\begin{eqnarray}}
  \newcommand {\eeqn} {\end{eqnarray}}

\begin{document}
\begin{titlepage}
\begin{flushright}{FTPI-MINN-05/02\\
UMN-TH-2343/05\\
NSF-KITP-05-03\\
hep-ph/0501200}

\end{flushright}
\vskip 0.8cm

\centerline{{\Large \bf Comments on Diquarks, Strong Binding and}} 

\vspace{2mm}

\centerline{{\Large \bf  a Large Hidden QCD Scale}}
\vskip 0.8cm
\centerline{\large M. Shifman and A. Vainshtein}

\vskip 0.4cm

\centerline{\em William I. Fine Theoretical Physics Institute, 
University of Minnesota,}
\centerline{\it Minneapolis, MN 55455}
\centerline{and}
\centerline{\em Kavli Institute for Theoretical Physics,
 UCSB, Santa Barbara, CA 93106}

\vskip 0.7cm
 
\begin{abstract}

We present arguments regarding the possible role of diquarks in low-energy
had\-ron phenomenology that has so far  escaped theorists' attention.
Good diquarks, i.e. the $0^{+}$ states of two quarks, are argued to
have a two-component structure with one of the components peaking 
at distances several times 
shorter than a typical hadron size (a short-range core). 
This can play a role in solving two old
puzzles of the 't~Hooft $1/N$ expansion:
strong quark-mass dependence of  vacuum energy density
and strong violations of the Okubo--Zweig--Iizuka (OZI) rule in 
quark-antiquark $0^\pm$ channels. 
In both cases  empiric data defy 't~Hooft's $1/N$  suppression.
If good diquarks play a role at an intermediate energy scale they
ruin 't~Hooft's planarity because of their mixed-flavor composition.
This new scale associated with good diquarks may
be related to a numerically large scale discovered in 
[V.~Novikov, M.~Shifman, A.~Vainshtein and V.~Zakharov,
Nucl.\ Phys.\ B {\bf 191}, 301 (1981)]
in a number of phenomena mostly related to vacuum quantum numbers 
and $0^\pm$ glueball channels.

If SU(3)$_{\rm color}$ of  bona fide QCD is replaced
by SU(2)$_{\rm color}$,  diquarks become 
well-defined gauge-invariant objects. Moreover, there is an exact symmetry 
relating them to pions. In this limit predictions regarding good diquarks
are iron-clad. If passage  from SU(2)$_{\rm color}$ to SU(3)$_{\rm color}$
does not lead to dramatic disturbances, these predictions 
remain qualitatively
valid in  bona fide QCD.
\end{abstract}

\thispagestyle{empty}
\end{titlepage}
\newpage

\section{Introduction}
\label{intro}

The constituent quark model enjoyed a remarkable success
in qualitative and semiquantitative descriptions of a huge body of data
on traditional hadron spectroscopy, static parameters
and other regularities
(for reviews, see e.g. \cite{Kok,1}). At the same time,  since the advent of QCD it has been 
known that some important hadronic
phenomena cannot be easily understood in this model.
Probably, the most clear-cut example of this type is 
the masslessness of pion (in the chiral limit), which can only
occur due to  a ``superstrong" attraction in the $0^-$ quark-antiquark channel. 
In the 1980's, a similar strong attraction was argued
to exist in the $0^+$ quark-quark channel.
This conjecture was supported by  occasional
observations of a special role played by diquarks in low-energy
hadronic phenomenology.

Indications of  diquark relevance, mentioned above,
came (i) from the instanton side \cite{S1} (see also \cite{NN}).
It was argued \cite{S1} that an instanton-induced interaction
is sufficient to form a bound antitriplet scalar diquark.
Then, (ii) there were suggestions \cite{NNN} (see also \cite{NN}) that  diquark correlations
are important in the enhancement of the operator $O_1$,
which contributes to the $\Delta I =1/2$ strangeness-changing decays.
In fact, an 
enhancement of matrix elements of $O_{1}$ for the hyperon decays
was first advocated 
in Ref.~\cite{SVZ} where this operator, together with penguin ones, 
was used to calculate all $\Delta I =1/2$ amplitudes. The penguin operators are
enhanced by a strong correlation in the quark-antiquark channel. 
The diquark representation of the operator $O_1$
introduced in Ref.~\cite{SVZ}, (see Eqs.\ (50) through (55))
was used to fix  sign
of the $O_1$ contribution in the hyperon decay amplitudes
and roughly estimate its absolute value. The sign was in agreement with
experimental data for all $S$ waves. 

Recently, the issue was revitalized by experimental evidence for existence of exotic pentaquark baryon (see \cite{Dzierba:2004db} for  recent reviews), the state which was predicted to be narrow in 
Ref. \cite{Diakonov:1997mm}   based on the chiral-quark-soliton model. Diquarks were  suggested as  an alternative explanation for pentaquark in a number of papers \cite{KL,W,WW,Shuryak:2003zi,Exotica,SW} which also discussed other  phenomenological evidences. The implications
are that  diquark correlations seem to be instrumental in  
excited and exotic hadron spectroscopy and in  other  aspects of hadronic physics,
and that ``good" diquarks in color-flavor locked antisymmetric combination, due to a strong attraction, are probably bound
to the extent that the ``mass" of a good diquark roughly coincides
with that of a constituent quark. 

It is natural to suggest that the characteristic size of the good diquark as well as that of the constituent quark is considerably  smaller than the nucleon size.  In the case of the constituent quark a relevant momentum scale in 
the $ \bar q q$  channel  is of order of a few GeV. We will argue that a similar scale appears in the diquark  $ q q$ channel.  A basic tool here is the observation
that if SU(3)$_{\rm color}$ is replaced by  SU(2)$_{\rm color}$,
diquarks become well-defined gauge-invariant objects, related
by  symmetry to  conventional Goldstone bosons (pions).

The existence of  hierarchical scales in QCD was discussed 
long ago in \cite{NSVZ}. The largest of these scales shows up in  $0^{\pm}$ glueball channels. 
Recent striking (and unexpected)
experimental evidence of a nonperturbative momentum scale much higher than $\Lambda_{\rm QCD}$ comes from  heavy ion collisions at  Relativistic Heavy Ion Collider (RHIC).  There, instead of weakly coupled quark-gluon plasma at short distances, the pattern of  hadron production shows hydrodynamical features and signals production of strongly interacting objects.

The existence of a hidden high scale could be helpful in understanding constituent quarks 
as small objects. 
In the diquark case a higher momentum scale implies a 
corelike structure similar to that of a pion.
This opens up the possibility of understanding the small width of a pentaquark (if its experimental status is confirmed). 
In this paper we use the idea of hierarchical scales to analyze diquark features and related phenomenological signals. We also present
arguments for special role of diquarks based on 
an expansion complementary to 't Hooft's large-$N_{c}$ limit ---
the so-called  ``orientifold'' large-$N$ expansion \cite{armoniN}.

It should be stressed that the introduction of diquarks in low-energy
hadron phenomenology is not meant to replace the 
quark model \cite{Kok}.
The old idea
that baryons are built of three (relatively weakly bound)
constituent quarks while mesons are similar composites
built of quark-antiquark pairs gives a reasonable overall picture of the hadronic world. It serves as a motivation for the SU(6)-based
description which produces a large number of results compatible with experiment. 
For instance, the ratio of  proton-to-neutron magnetic moments,  $-1.47$, is very close to $-3/2$  as predicted  in
the quark model.
The SU(6)-based prediction of the ratio $\pi$-to-$\rho$ charge radii
is also consistent with what is known from experiment.
The reader can easily continue this~list.

The diquark short-range correlations we discuss below are 
meant to supplement the quark model
by explaining a few subtle aspects of hadron phenomenology
which do not come out right in the quark model.
If we had a fully developed dynamical scheme
of the quark-gluon low-energy interactions
we would first confirm that this scheme
is compatible with the quark model in all cases
(the majority!) where the latter produces successful
predictions, while diquarks show up only
in certain aspects (they were mentioned above)
where the quark model fails. Alas ... no fully developed 
dynamical model of this type is available at present.
Consideration of diquarks in the present paper
 is carried out largely
at the qualitative rather than quantitative level.
In this sense our results are by no means ``carved in stone."
Our main intention is to provoke further discussion of the issue
and related activities.

\section{Diquarks' progenitors}

The usefulness of diquark notion in hadronic physics
is based on the assumption of an ``abnormally strong"
attraction in the $0^+$ channel. Moreover, it is meaningful
only if there are two scales separated by a rather large numerical factor:
the size of a good diquark $R_{dq}$ and that of a typical hadron $R_h$,
which may be larger, say, by a factor of $\sim 3$. 
 (The smallness of the ratio
$R_{dq}/R_h$ is not parametric, but, rather, of a numerical nature.)
Strongly bound diquarks must have a smaller typical size
than that of typical hadrons. Otherwise the picture 
advocated in Refs.~\cite{KL,W,WW,Shuryak:2003zi,Exotica,SW} 
could hardly be consistent.

Let us ask ourselves if we are familiar with other objects
with similar dynamics. The answer is yes. 
In the $0^-$ $\bar q q$ channel a strong attraction between
the quark and antiquark leads, in the chiral limit, to formation
of massless pions.
One can give a strong argument for the existence of
a short-range core in the pion.

Indeed, let us compare two matrix elements,
\begin{equation}\label{me}
\begin{split}
\langle 0|\bar d \gamma_{\mu} \gamma_{5}u|\pi^{+}\rangle &=if_{\pi}\,q_{\mu}\,,
\\[1mm]
\langle 0|\bar d  \gamma_{5}u|\pi^{+}\rangle &=if_{\pi}\,\frac{m_{\pi}^{2}}{m_{u}+m_{d}}\,.
\end{split}
\end{equation}
The second matrix element is enhanced by the ratio
\beq\label{hidscale}
\frac{m_{\pi}^{2}}{m_{u}+m_{d}}\approx 1.8~{\rm GeV}\,,
\eeq
which has a smooth chiral limit  {$m_{q}\to 0$} but is
rather large numerically, because of $m_{\pi}/(m_{u}+m_{d})\approx 13$.

\begin{figure}[h]
\centerline{\includegraphics[width=2.7in]{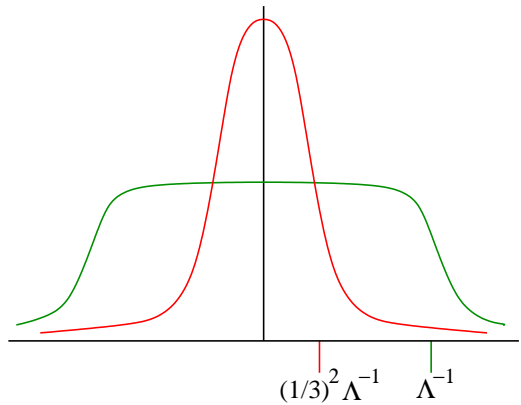}}
\vskip -18mm
\caption{\small Double-component structure with a core.} 
\label{fig:twocompon}
\end{figure}

The {$1^{+}$}  and {$0^{-}$} currents, $\bar d \gamma_{\mu} \gamma_{5}u$
and  $\bar d  \gamma_{5}u$, look very different in 
the framework of QCD sum rules \cite{SR}.
The first one works well and allows one to calculate $f_{\pi}$, 
while the second 
essentially does not work with perturbative Operator Product  Expansion (OPE) coefficients. 
In other words, the intervals of duality are quite different,
\begin{equation}
\begin{split}
\bar d \gamma_{\mu} \gamma_{5}u\,:\qquad 0.6~{\rm GeV}^{2},\\[1mm]
\bar d  \gamma_{5}u\,: \qquad \quad 1.9~{\rm GeV}^{2}.
\end{split}
\end{equation}
How can one interpret this fact? In the quantum-mechanical approach 
the matrix elements (\ref{me}) can be viewed 
as a value of wave function at zero separation, {$\psi(0)$}.
We have two of them: the leading twist, $t=2$ for {$\bar d \gamma_{\mu} \gamma_{5}u$} 
and nonleading twist, $t=3$, for {$\bar d \gamma_{5}u$}, 
\begin{equation}
\psi=\alpha_{2}\psi_{t=2}+\alpha_{3}\psi_{t=3}\,.
\end{equation}
The large value of {$\psi_{t=3}(0)$} for  the {$\bar d \gamma_{5}u$} component 
implies the existence of  a smaller size. 
We try to illustrate this in Fig.\,\ref{fig:twocompon}.
The factor $(1/3)^{2}$ for core size will be interpreted below in the framework 
of the instanton liquid model. 
Of course, the picture is rather symbolic, not only characteristic sizes
but also relative probability amplitudes (coefficients $\alpha_{2,3}$) influence 
results for different probes. Moreover, the profile of
wave function of the leading twist should also involve a shorter scale. This follows from 
a smallness of $f_{\pi}\simeq 135$~MeV in comparison to the characteristic momentum scale 
$m_{\rho}\simeq 776$~MeV; note that  $f_{\pi}\propto \sqrt{N_{c}}$ and grows with the number of colors while $m_{\rho}$ is stable in this limit.

The instanton liquid model \cite{ILM} provides a particular way of interpreting a hierarchy of scales.
The model operates with two parameters, the average instanton size 
$\rho=0.48\,\Lambda_{\rm QCD}^{-1}$ 
which is a factor of $\sim 3$ smaller than the average instanton separation $R=1.35\, \Lambda_{\rm QCD}^{-1}$.
Instantons are Euclidean objects and to relate their parameters with our Minkowski world, note that
$R^{-1}\sim \Lambda_{\rm QCD} $ is a typical hadronic scale while $\rho^{-1}$ is the geometrical mean
between $\Lambda_{\rm QCD}$ and the higher glueball scale 
$\Lambda_{\rm gl}$.   One can readily verify this at weak coupling in gauge theory with the adjoint Higgs field,
where the inverse instanton size is the geometrical mean between $W$ boson mass and that of monopoles (or spalerons for the Higgs field in the fundamental representation).

Hence, we conclude that in the glueball world
$\Lambda_{\rm gl}\sim 3^{2} \Lambda_{\rm QCD}$, which explains $(1/3)^{2}$ in Fig.\,\ref{fig:twocompon}. This is consistent with the estimate in \cite{NSVZ}
based on the low-energy theorem for the correlator of $G_{\mu\nu}G^{\mu\nu}$.
Let us add that in the instanton liquid model the smallness of $f_{\pi}$ appears as 
$f_{\pi}\sim \rho / R^{2}\sim \Lambda_{\rm QCD}(\Lambda_{\rm QCD}/\Lambda_{\rm gl})^{1/2}$, while the large scale \eqref{hidscale} is $m_{\pi}^{2}/(m_{u}+m_{d})\sim  R^{2}/\rho^{3}\sim \Lambda_{\rm QCD}(\Lambda_{\rm gl}/\Lambda_{\rm QCD})^{3/2}$.

If pions serve as a textbook dynamical example of a strongly bound
quark-antiquark system, what can be said  of diquarks?
Of course, in  discussing diquarks,  it is rather difficult to stay 
fully quantitative
since they are not defined as gauge-invariant objects in QCD.
Is there a limit in which pions and diquarks are
related by a rigorous symmetry?

Such a limit does exist with the assumption that
replacing the SU(3) color group of QCD
by SU(2) we do not impose a dramatic change
on hadron dynamics (for a more precise statement see below).
In two-flavor SU(2)$_{\rm color}$  QCD with massless quarks the flavor group
is enhanced to SU(4), and the pattern of its
spontaneous symmetry breaking is different
from  bona fide QCD \cite{dp,KSV}. The reason is that in SU(2)
the fundamental representation is (quasi)real, i.e. antiquarks form 
the same representations as quarks and interact with gluons the same way, too.
Two flavors are composed of four two-component Weyl spinors
(left-handed) and four complex conjugated  ones (right-handed).
In \cite{KSV} it was shown, by 't Hooft's matching \cite {thooftch}
of triangles, that the pattern of spontaneous breaking of flavor
symmetry is as follows:
\beq
\mbox{SU(4)} \to \mbox{SO(5)}\,.
\label{pat}
\eeq
This pattern was confirmed by other methods 
(for a recent review see \cite{Ver}), and used to analyze different phases of the theory depending on chemical potential and temperature \cite{Rapp:1997zu,Klein:2004hv}. The partnership of pions and diquarks in SU(2)$_{\rm color}$ was also discussed in Ref. \cite{DiakPet}.

To elucidate (\ref{pat}) let us note
that  the flavor doublets of the left-handed quarks $(u_{ L \alpha},\;
d_{ L \alpha})$ and antiquarks $(\bar d_{ R \alpha},\; -\bar u_{ R\alpha})$ ($\alpha=1,2$) form 
the fundamental representation of SU(4) 
\beq
\chi^{i}=\left(\!\!\!\!\begin{array}{r}u_{L}\\d_{L}\\ \bar d_{R}\\-\!\bar u_{R}\end{array}\right),
\eeq
containing entries with different baryon numbers. 
The six bilinears $\chi^{i}\chi^{j}$ (convolution in space and color
indices is implied) antisymmetric in $i,j$ can be viewed as an O(6) vector.
A nonvanishing vacuum average of this vector which 
can be aligned as 
\beq
\langle \bar u u + \bar d d \rangle = \langle \bar u_{R} u_{L} +\bar u_{L} u_{R}  + \bar d_{R} d_{L}
+  \bar d_{L} d_{R}\rangle\neq 0 
\eeq
implies the spontaneous symmetry breaking
which preserves SO(5). 
Acting on $\bar u u + \bar d d$ by SU(4) generators we determine five Goldstone bosons  
corresponding to operators
\begin{subequations}
\begin{eqnarray}
&&\bar u_{R}u_{L}-\bar d_{R}d_{L}-(R\leftrightarrow L)= -\bar u \gamma_{5}u +\bar d \gamma_{5}d\,,
\nonumber \\[1mm]
&&\bar u_{R}d_{L}-(R\leftrightarrow L)= -\bar u \gamma_{5}d \,,\qquad
~~\bar d_{R}u_{L}-(R\leftrightarrow L)= -\bar d \gamma_{5}u \,,
 \\[2mm]
&& u_{L}d_{L}+(R\leftrightarrow L)=-uC\gamma_{5}d\,, \qquad 
\bar d_{L}\bar u_{L}+(R\leftrightarrow L)=-\bar dC\gamma_{5}\bar u\,.~~~~~~
\label{udscal}
\end{eqnarray}
\end{subequations}
Here $C$ is the charge conjugation matrix.
In addition to the
triplet of massless $0^{-}$ pions we get two ``baryon'' 
Goldstone bosons which are 
$0^{+}$ diquark states.

In terms of symmetry currents, 
fifteen currents generating SU(4) are classified as
ten currents which are generators of the linearly realized SO(5), 
plus
five spontaneously broken currents, which have the form
\begin{subequations}
\beqn
&&\bar u\gamma_\mu\gamma_5 d\,, \quad 
\bar d\gamma_\mu\gamma_5 u\,,
\quad \bar u\gamma_\mu\gamma_5 u - \bar d\gamma_\mu\gamma_5 d\,,
\\[1mm]
&&u C\gamma_\mu\gamma_5 d\,,\quad  \bar u\gamma_\mu\gamma_5 C\bar d\,,
\label{udcurr}
\eeqn
\end{subequations}
and are coupled to the Goldstones.

Thus, we have more than the conventional triplet of pions;
there emerge two extra diquark states related to pions by exact symmetry.
Consequently, these diquark states also have the two-component
structure reflecting the existence of a higher scale.

The question is, what happens to these states as we
elevate the gauge group from SU(2)$_{\rm color}$ to SU(3)$_{\rm color}$?
It is not difficult to see that the extra states become diquarks.
This passage from SU(2)$_{\rm color}$ to SU(3)$_{\rm color}$  can be formalized if we introduce
into SU(3)$_{\rm color}$ theory an additional complex scalar field
(``Higgs") in the triplet representation, and analyze the theory as
a function of the vacuum expectation value (VEV)
of this field. It is well-known
\cite{fradshen}
that there is no phase transition in VEV -- physics smoothly flow from the SU(2)$_{\rm color}$
phase at large VEV to SU(3)$_{\rm color}$ phase at small (vanishing) VEV
of fundamental Higgs.  

If $N_f >2$ (and the color group is SU(2)$_{\rm color}$) the flavor symmetry of the theory is SU($2N_f$). The pattern of spontaneous chiral symmetry breaking is \cite{KSV,Ver}
$$ {\rm SU}(2N_f) \to Sp (2N_f) \,$$
(note that $Sp(4)$ is the same as SO(5)). The linearly realized part of symmetry, $Sp (2N_f)$, has $2N_f^2 + N_f$ generators.
There are $2N_f^2 - N_f-1$ Goldstone states.
Out of those,   $N_f^2  -1$  are conventional pions.
The remaining  $N_f^2  -N_f$ Goldstone states are ``baryonic."
Upon lifting SU(2)$_{\rm color}$ to SU(3)$_{\rm color}$
these $N_f^2  -N_f$ baryonic Goldstones become diquarks.

Since in SU(2)$_{\rm color}$ theory
baryonic Goldstones and pions are related
by an exact symmetry, their spatial structure is the same.
In particular, the two-component structure depicted in Fig. 1
is shared by both. As we pass to SU(3)$_{\rm color}$,
the exact symmetry no longer holds, but an approximate similarity
in the spatial structure of pions and diquarks is expected to hold.

The symmetry between pions and diquarks
in  SU(2)$_{\rm color}$ theory gives us  an idea of the momentum
interval in which composite diquarks might play a role.
If pion loops make any sense, this can only happen at
$130\,{\rm MeV} \lesssim p \lesssim 700\, {\rm MeV}$. The same
must be  applicable to diquark loops.

In {\em bona fide} QCD, with SU(3)$_{\rm color}$, in
the energy interval between $\sim R_h^{-1}$ and
$\sim R_{dq}^{-1} $ good diquarks act as pointlike 
color-antitriplet objects
whose interaction with gluons is determined only by
the color representation to which they belong, in much in the same way
as color-triplet quarks.

\section{Approaching from the large-\!\boldmath{$N$} side}

Everybody knows that 't Hooft's $1/N$ expansion \cite{thooft}
presents a powerful tool in qualitative and semiquantitative analysis of QCD, and as an organizational principle. A complementary ``orientifold"
large-$N$ expansion
 was suggested recently \cite{armoniN} (see also
\cite{corramond}). 
This expansion is similar in spirit --- but not technically
--- to expansion which goes under the name of 
 {\em topological expansion} and was suggested long ago \cite{TE}.
(Topological expansion assumes that the number of flavors
$N_f$ scales as $N$ in the large-$N$ limit, so that the ratio
$N_f/N$ is kept fixed. 
A quark-diquark representation of nucleon is natural in topological expansion.)

The orientifold 
large-$N$ extrapolation treats quarks as Dirac fields in 
the two-index antisymmetric representation of the SU($N$) color group.
At $N=3$ the two-index antisymmetric representation is equivalent to fundamental; hence,  the starting point for both extrapolations --- orientifold and 't Hooft's --- is  bona fide $N=3$ QCD, and both are equally suitable for $1/N$ expansion.

In most instances orientifold and 't Hooft large-$N$ expansions
lead to overlapping predictions. There are notable exceptions, however,
of which the most important are:
\begin{itemize}
\item[]
(i) quark-mass dependence of vacuum energy density, and 
\item[]
(ii) the OZI rule. 
\end{itemize}
In the former case, according to 't Hooft, quark-mass
dependence should be suppressed by a factor $1/N$
as all quark loops scale as $1/N$ at large $N$. 
On the other hand, in the orientifold theories quark loops
are {\em not} suppressed, and one should expect quark-mass dependence
of a ``natural" order of magnitude (see below).
The very same property, $1/N$ for each  ``additional"  quark loop in 
the 't Hooft limit, is supposed to explain inhibition of transitions
between quarks of distinct flavors (the OZI rule). 
In the orientifold large-$N$
limit  quark loops do not carry $1/N$, and, correspondingly,
transitions
between quarks of distinct flavors are not suppressed by $1/N$.

At large $N$ 't Hooft and orientifold extrapolations
present distinct theories, and, therefore, it is not surprising  
that some predictions do not coincide. There is one point,
$N=3$, where these theories do coincide, and, hence, must lead 
to the same results. Logically there are two options that would lead
to a reconciliation at $N=3$:
a numerical suppression of appropriate amplitudes in orientifold theory 
or a numerical enhancement in  't Hooft's  theory. We will argue that
it is the latter option that is realized in certain important cases
(see below).

On the phenomenological side, both points
(i) and (ii) above, rather than being successful,  present a serious challenge to the 't Hooft approach. Indeed, two fundamental relations
\beq
\frac{d}{dm_q} {\cal E}_{\rm vac} =\langle \bar q\, q\rangle 
\eeq
and
\beq
{\cal E}_{\rm vac}\Big |_{m_{q}=0} =
\frac{\beta(\alpha_{s})}{16 \alpha_{s}}
\left\langle G_{\mu\nu}^a\,G^{\mu\nu\, a }\right\rangle
\eeq
being combined and evaluated numerically \cite{NSVZ}
imply that in QCD with light quarks the vacuum energy density 
${\cal E}_{\rm vac}$ changes by a factor of $\sim 2$ as the strange quark mass increases from zero to its actual value $\sim 150$ MeV.
Moreover, while the extra quark-loop suppression inherent in the
't Hooft $N$ counting seemingly does explain the OZI rule
in the $1^-$ channel, where the $\phi$ meson is an almost pure
$\bar s s$ state, it badly fails in $0^\pm$ channels where 
flavor mixing is fully operative:
say, the  composition of $\eta^\prime$ meson only slightly
deviates from $\bar uu+\bar dd+ \bar s s$.

It was suggested \cite{NSVZ} that the 't Hooft-defying enhancement in
these two cases is due to ``direct" instantons. Instantons are theoretical objects existing in 
 Euclidean space-time. It is natural to ask what 
actual physics lies behind this phenomenon. What is the relevant 
Minkowski-space picture? As was mentioned in Sect. 2, simultaneously, 
the existence of a new numerically large scale in hadronic physics
was discovered in \cite{NSVZ}. In phenomena where
vacuum quantum numbers (e.g. the $0^\pm$ channels)
were involved, a hidden energy scale,
larger than  the conventional
$\sim 300$ MeV  by a factor of $\sim 10$ or so,   was revealed.

The space-time structure of good diquarks is   (in  Weyl notation):
$\chi^{[ if}_\alpha \, \chi^{jg]\, \alpha}$
where square brackets denote antisymmetrization with respect
to both color indices $i,\,j$ and flavor indices
$f,\,g$; and $\alpha $ is the spinor index. Thus, they are bosons.
Good diquarks are scalars. The flavor mixing is maximal in good diquarks. 
They appear, right from the start, as
two-index antisymmetric  objects in color representation. 
This naturally matches them with orientifold theory. 

If $N$ is treated as a free parameter, then
diquarks  propagating  
in loops with virtual momenta in the interval
$(\sim R_h^{-1}\,,\,\, \sim R_{dq}^{-1}) $
have no $1/N$ suppression. This simple observation suggests that
they play a crucial role in strong quark-mass dependence of vacuum energy
as well as  in the failure of the OZI rule in $0^\pm$ channels,
acting as Minkowski-space ``representatives"   of the
Euclidean instantons much in the same way as sphalerons and monopoles are Minkowski signatures of instantons  at weak coupling. 
Note that, being composite in flavor,
diquark loops (with diquarks treated as
pointlike objects) also negate the notion of planarity.

At $N=3$ the failure of the OZI rule in $0^\pm$ channels
should look like a numerical enhancement with regards to the
't Hooft counting. For instance, the 
VEV $\langle \bar s s\rangle $ is contributed
by  loops of $(su) $ and $(sd)$ diquarks;
both can be scalar and pseudoscalar (see Fig. 2).
\begin{figure}[h]
\centerline{\includegraphics[width=1.5in]{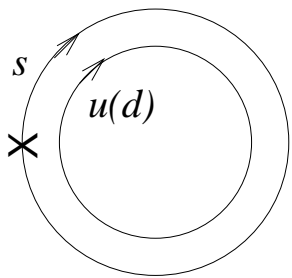}}
\caption{\small%
The diquark contribution to $\langle\bar s s \rangle$.}
\label{diqu-one}
\end{figure}
This provides a numerical enhancement factor $\sim 4$.
Moreover, if this picture is correct,
it is absolutely obvious 
that in $0^+$ channels at low energies
flavor mixing has no suppression whatsoever
(the same is true for $0^-$). An immediate consequence 
is that the correlation functions
\beq
\langle \bar q(x) q(x)\,,\, \bar q(y) q(y) \rangle\, ,\qquad
\langle \bar q(x)\gamma^5 q(x)\,,\,\bar q(y) \gamma^5 q(y) \rangle\,
\eeq
in addition to two-quark mesons in the intermediate state,
must also exhibit four-quark exotic mesons.

Why then does flavor transition inhibition work so well in $1^-$
channels ($\varphi$ meson decays)? The answer may be 
purely kinematical.
In the $J=1$ case the diquark pair must be in the $P$ wave
which  increases effective energies to the
extent that diquarks no longer can be viewed
as local; they 
are essentially destroyed and do not contribute in loops.
The destruction of diquarks resurrects planarity. 
In orientifold large-$N$ expansion
the OZI rule in $1^-$
channels is  explained by
relatively large effective energies and numerical smallness of
corresponding loop factors which become
quite suppressing (two loops!).

\section{Diquarks and weak nonleptonic decays of heavy baryons}

As was mentioned in Sect. \ref{intro}, the short-distance diquark component
is presumably responsible for (a part of)  the enhancement of nonleptonic decays 
of strange 
hyperons.\footnote{\,In addition to diquark correlations,
 the enhancement is also due to the
quark-antiquark correlations, i.e. the penguin mechanism which, simultaneously,
explains the enhancement 
in  $K$ meson decays \cite{SVZ}.} 
It is natural to ask whether a similar enhancement occurs 
for baryons  containing heavy $c$ and $b$ quarks. 

Let us start from $b$-containing baryons. 
The phenomenological issue here is that diquarks could play a role
in reconciling experimental trends with theoretical expectations
in the problem of lifetimes of $b$-containing hadrons,
in particular, $\Lambda_b$. This issue has a long and dramatic
history. 
Asymptotically (at $m_b\to\infty$)
lifetimes of all  $b$-containing hadrons must be equal.
At finite but large $m_b$ deviations of the ratio
$\tau (\Lambda_b)/\tau (B_d) $ from unity can be
calculated as an expansion in   powers of $1/m_b$.
A (formally) leading power correction $m_b^{-2}$
is due to the dimension-5  chromomagnetic
operator  parameterized by $\mu_G^2$. It contributes at the level
$\sim\! - 0.02$, see e.g. \cite{Bigi-one,Bigi-two,Bigi-three}
and is rather unimportant. 
Moreover,  this correction is flavor blind. Flavor-dependent and 
more important numerically
are dimension-6 four-quark operators whose contribution is suppressed by
$m_b^{-3}$.
The original estimates \cite{Shifman-one} of four-quark operators
in heavy baryon lifetimes were at the level $\sim\! - 0.02$.

At the same time, experimental measurements conducted in the
1990's seemingly indicated that the ratio $\tau (\Lambda_b)/\tau (B_d)$
could be as small as $\sim 0.8$, i.e. preasymptotic corrections 
as large as $\sim - 0.2$. This apparent discrepancy
ignited a renewed theoretical effort and extensive debate in the literature 
\cite{Uraltsev-one,Voloshin,APetrov,Sachrajda} (the last paper reported
lattice results). It is worth emphasizing that
there are two distinct mechanisms responsible for
$m_b^{-3}$ corrections in $\tau (\Lambda_b)/\tau (B_d)$,
namely,  {\em Pauli interference} and {\em weak scattering}.
Pauli interference works in the ``unfavorable"
direction, increasing the ratio $\tau (\Lambda_b)/\tau (B_d)$,
while weak scattering tends to decrease it.
Stretching estimates of the matrix elements
of relevant four-quark operators in the ``favorable"
direction one typically gets $\sim\! 0.03$ and $\sim\! - 0.07$
for  Pauli interference and  weak scattering, respectively,
so that there is  significant compensation
\cite{Uraltsev-one}.

\begin{figure}[h]
\begin{center}
\begin{minipage}[h]{12cm}

\centerline{\includegraphics[width=8cm]{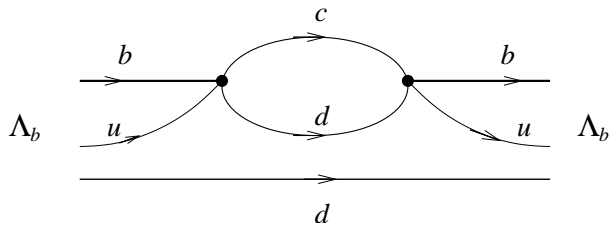}}

\caption{\small%
Preasymptotic $m_b^{-3}$ corrections: the 
weak scattering mecha\-nism
giving rise to the four-quark operator
(\ref{lbdq}) in the  OPE for $\tau (\Lambda_b)$. }

\label{weaks}

\end{minipage}
\end{center}
\end{figure}

For our purposes it is important to note that the four-quark operators emerging in
 Pauli interference and  weak scattering 
have spatial structures which do  not coincide for these two
mechanisms. In weak scattering the only relevant operator
occurring in OPE (Fig.~\ref{weaks}) has a structure
``good diquark density times good diquark density,"
of the type mentioned for $O_1$ in Ref. \cite{SVZ}, namely,\footnote{%
\,Strictly speaking $j_{k}$ represents a combination of the
$0^{+}$ and $0^{-}$ diquarks.}
\beq
O_- = 2 \left( j_{k}\right)^\dagger  \left( j_{k }\right) ,
\qquad j_{k } = \varepsilon_{kji}\, {b^{j}}C \,\frac{1-\gamma_5}{2}\,
u^{i}\,,
\label{lbdq}
\eeq
where $i,j,k$ are color indices.
The assumption that a strong (positive) diquark
correlation persists in the $0^+$ system of one heavy and one light
quark, would result in a  further enhancement of
the weak scattering contribution in $\Lambda_b$,
effectively destroying the cancellation between
Pauli interference and  weak scattering. In this case
theoretical prediction $\tau (\Lambda_b)/\tau (B_d) \sim 0.9$
would become natural. 

Note that recent experimental measurements
of the ratio $\tau (\Lambda_b)/\tau (B_d) $ tend to shift the central value
of the ratio from $\sim\!\!0.8$ up to $\sim\!\!0.9$. Note also that 
the argumentation above 
is applicable  to $\Xi_{b}$ baryons as well. 

One can give another argument in favor of a
certain enhancement of the $\Lambda_b$ matrix element
of the operator (\ref{lbdq}), compared to the naive estimates of
the 1980's. This argument goes
along the lines discussed in Sect.\ 2.
There, reducing the color group from SU(3) to SU(2),
we were able to relate $0^+$ light diquarks to pions.
If we follow the same strategy here, we will relate
the $(bu)$ good diquark to $B$ mesons. Extensive analyses in the 1990's firmly 
established the fact that $f_B$ is quite large, $f_B\sim 200$
MeV, a factor of $\sim\!1.5$ larger 
than  was previously believed.
In SU(2)$_{\rm color}$
the factor of 2 enhancement of $f_B^2$
immediately translates into a similar enhancement of
 $\langle \Lambda_b | O_- | \Lambda_b\rangle $.

Now let us briefly comment on the situation with $c$-containing baryons.
This case is marginal: on the one hand the charmed quark can be viewed as heavy,
and heavy quark-mass expansion could be applied. As seen from
explicit calculations \cite{Shifman-one,Blok}, this expansion is rather
poorly convergent and can be used only for qualitative estimates. 
This fact indicates that the $c$ quark is not heavy enough. 
The hidden scale quoted in Eq.\,\eqref{hidscale}
is actually of the same order as $m_{c}$. 

In this marginal situation one can try the opposite description, treating 
the $c$ quark as ``light."
The $cd$ diquark plays a similar role in $c$-baryon nonleptonic 
decays as the $ud$ 
diquark in hyperon decays. What is different is that the penguin 
mechanism plays no significant role in total widths of 
$c$-containing hadrons. That is why one can qualitatively 
expect a relative enhancement of all $c$-baryon decays over meson ones. This expectation is indeed confirmed by experimental data.

\section{Hidden scale and deep inelastic processes}

Deep inelastic scattering (DIS) provides a good probe for 
diquark correlations in nucleons. At first sight
an immediate consequence of these 
correlations is an enhancement of the higher-twist 
correction which carries an extra power of $1/Q^{2}$, namely $Q^{(-t+2)}$.  
For local operators twist $t$ is the difference between 
its dimension and spin.
The nucleon average of
higher-twist operators containing the diquark 
combinations of fermion fields becomes 
large if the corelike diquark structure is present. 

It is simple, however, to demonstrate \cite{VJ} that the effect does 
not show up at the
twist-4 (next-to-leading) level. The leading level is twist $t=2$. 
Indeed,
to generate a twist-4  effect the corresponding
operator should contain diquark combinations of fermion fields
in a form similar to $j_{k}$ in Eq.\,(\ref{lbdq}),
\begin{equation}
\epsilon^{\alpha\beta}\epsilon_{ijk}\epsilon_{fgh}q_{L\alpha}^{if}q_{L\beta}^{jg}\,, \qquad
\epsilon^{\dot\alpha\dot\beta}\epsilon_{ijk}\epsilon_{fgh}q_{R\dot\alpha}^{if}q_{R\dot\beta}^{jg}\,,
\label{comb}
\end{equation}
where $\alpha,\beta,\dot\alpha,\dot\beta=1,2$ are Weyl spinor indices, $i,j,k=1,2,3$ and $f,g,h=1,2,3$
refer to quark colors and flavors.
The twist of these combinations  equals 3 (it remains
intact if some number 
of covariant derivatives $D_{\mu}$ are 
inserted). For DIS we consider operators with vanishing
baryon charge; therefore,  diquark combinations 
(\ref{comb}) must be multiplied by operators 
$\bar q D\ldots D \bar q$. The minimal twist 
of these operators is 2, so the total twist is not less than 5. 
Actually the minimal total twist is 6. Twist 5 is excluded because 
 of the chiral features of DIS operators under 
SU(3)$_{L} \times$SU(3)$_{R}$ transformations. 

Thus, diquark correlations do  not show up at the 
level of $1/Q^{2}$ corrections. This is  good
because no enhancement in these corrections was observed 
experimentally. What could  remain   
is an enhancement for twist 6, i.e. in the $1/Q^{4}$ corrections. 

On the other hand, once we consider the range  of 
moderate momentum transfers $Q$, smaller than the hidden scale,
we can view corelike diquarks as pointlike spin-zero ``quarks." This implies a strong impact on the longitudinal structure functions which vanish for spin 1/2 partons but not for spin 0. 
What do experimental data show for longitudinal cross sections 
$\sigma_{L}$ in the range of relatively low 
$Q^{2}<3~{\rm GeV}^{2}$? 

A glance at the data 
for $\sigma_{L}$ \cite{Whitlow:1990gk,Yang:1999xg,data}  
shows that $\sigma_{L}$ maximizes at 
$Q^{2}\sim 2~{\rm GeV}^{2}$ (let us recall that $\sigma_{L}$ vanishes at $Q^{2}=0$ and at infinity).
The amplitude of the effect was first interpreted \cite{Whitlow:1990gk} as excess over 
predictions based on perturbative QCD (with target mass corrections accounted), 
i.e. as an indication of nonperturbative higher-twist effects. However, 
later  it was shown in Ref. \cite{Yang:1999xg} that inclusion of  next-to-next-to leading (NNLO) terms 
in the perturbative fit allows one to get rid of higher twists.  

We think that in the considered 
range of $Q^{2}$ untangling of NNLO corrections 
from nonpertubative effects is ambiguous. In perturbation theory the observed enhancement implies rather large higher-order corrections, of the order of the effect  itself. 
Such a situation could leave space for 
a significant role for nonperturbative effects.
Thus, the possibility of a diquark enhancement in $\sigma_{L}$ is not completely excluded.
A dedicated 
analysis of the moments of longitudinal 
structure functions at moderate $Q^{2}$ 
could shed light on the issue.

\section{Conclusions}

This paper could have been called 
``Connecting diquarks to pions,"
or ``Diquarks and a Large Scale,"
or ``Diquarks vs. $1/N$ Expansions." 
We discussed various approaches allowing one  to attempt
to quantify the role of diquarks in hadronic physics.
The most solid consideration, albeit somewhat remote from  bona fide
QCD, is that based on SU(2)$_{\rm color}$.
Reducing the gauge group from SU(3) to SU(2)
allows one to relate diquarks and pions through a global
symmetry which exists only for SU(2)$_{\rm color}$.
Diquarks become well-defined gauge-invariant objects,
which share with pions a two-component structure
with a relatively short-range core. 
Then one can speculate, qualitatively or, with luck, semiquantitatively
on what remains of this symmetry upon lifting SU(2)$_{\rm color}$
to SU(3)$_{\rm color}$. 
It is worth noting that all instanton-based calculations carry a strong
imprint of the above symmetry since basic instantons are, in essence, SU(2)$_{\rm color}$ objects.

In the case of pions a hidden large scale is known to exist based on various arguments.
Our task was to investigate consequences of the existence of this scale in diquarks. 
We argue that short-range diquark correlations
play an important role in QCD vacuum structure: they help resolve
the apparent contradiction with large-$N$ expectations for vacuum structure.
We also discuss consequences of diquarks in nonleptonic baryon decays.
In all cases --- hyperons, $c$- and $b$-containing baryons --- 
a short-range diquark core 
seems to be instrumental in understanding existing phenomenology.

Finally, we addressed the issue of diquarks in application to deep inelastic scattering.
In particular, they could show up in longitudinal structure functions at moderate $Q^{2}$.
The current experimental situation seems to 
be inconclusive and calls for  further analysis.

The issue of diquarks was raised and discussed mostly in application to hadron spectroscopy
in Refs. \cite{KL,W,WW,Shuryak:2003zi,Exotica,SW}. We extended the discussion by including for consideration a hidden large scale   implying a short-range core in diquarks. We tested the idea in a range of applications other  than hadronic spectroscopy. The idea survived these tests. However, our considerations are essentially qualitative; more quantitative analyses are most welcome. 
 
\section*{Acknowledgments}

We are grateful to I. Bigi, A. Capella, D. Diakonov, L.~Glozman, R.~L.~Jaffe, 
M.~Karliner, H.~J.~Lipkin, A. Shuryak, M.~Strikman, N. Uraltsev, 
M. Voloshin, and F.~Wilczek for helpful 
comments on the draft and useful discussions
and communications. We would like to to thank R. Keen for 
proofreading the manuscript.

This work was carried out, in part, within the framework of
the program {\sl QCD and String Theory} at the
Kavli Institute for Theoretical Physics,
UCSB, Santa Barbara, CA, ÊAugust 2 -- December 17, 2004.
We are grateful to colleagues and staff of KITP 
for kind hospitality, and acknowledge financial support of the National 
Science Foundation under Grant No. PHY99-07949.
This work    was also
supported in part by DOE grant DE-FG02-94ER408.


\end{document}